\documentclass[prl,twocolumn,aps]{revtex4}

\usepackage[isolatin]{inputenc}

\usepackage[dvips]{graphicx}

\begin{document}

\title{Stoner-Wohlfart model applied to bicrystal magnetoresistance hysteresis}

\author{R. Gunnarsson}
\email[]{robert.gunnarsson@mc2.chalmers.se}\affiliation{Quantum
Device Physics Laboratory, Department of Microtechnology and
Nanoscience, Chalmers University of Technology, SE-412 96
G\"oteborg, Sweden}

\author{M. Hanson}
\affiliation{Department of Experimental Physics, Chalmers
University of Technology and G\"oteborg University, SE-412 96
G\"oteborg, Sweden}

\author{C. Dubourdieu}
\affiliation{Laboratoire des Mat\'{e}riaux et du G\'{e}nie
Physique, CNRS UMR 5628, ENSPG BP46, 38402 St. Martin d'H\`{e}res,
France}

\begin{abstract}
We calculate numerically the magnetization direction as function
of magnetic field in the Stoner-Wohlfart theory and are able to
reproduce the shape of the low-field magnetoresistance hysteresis
observed in manganite grain boundary junctions. Moreover, we show
that it is necessary to include biaxial magnetocrystalline
anisotropy to fully describe the grain boundary magnetoresistance
in La$_{0.7}$Sr$_{0.3}$MnO$_3$/SrTiO$_3$ bicrystal tunnel
junctions.
\end{abstract}


\maketitle

Can the fairly simple model presented by Stoner and Wohlfart in
1948 \cite{Stoner1948} provide insights in the processes that
occur in manganite bicrystal grain boundary junctions? In the
following we show how the coherent rotation of the magnetization
vector can explain the characteristic features in the field
dependence of the magnetoresistance.

Recently, a substantial low-field magnetoresistance was observed
in grain boundaries of perovskite manganites
\cite{Hwang1996,Gupta1996} and other half-metallic ferromagnets
\cite{Hwang1997,Yin2000}. In order to understand and exploit this
effect several studies have used bicrystal grain boundary
junctions \cite{Mathur1997,Steenbeck1997,Gross2000}. Since
bicrystals are samples with well defined crystal orientations with
respect to an interface, they are in a way ideal systems for
studies of the behaviour of magnetic tunnel junctions.
Traditionally, the conductivity of magnetic tunnel junctions is
compared with Julliere's model \cite{Julliere1975}, in which the
tunneling magnetoresistance is determined by the spin polarization
of the electrodes. A more realistic scenario was considered by
Slonczewski \cite{Slonczewski1989}, who used a method to match the
wave functions across the tunneling barrier. He derived an
expression for direct tunneling through the interface which
includes the angle between the directions of the magnetization of
the electrodes.

However, direct tunneling is not sufficient to explain the
measured transport data in manganite grain boundary junctions. In
fact, even though several scenarios have been suggested to explain
measured data, the transport mechanism in bicrystal grain
boundaries of manganites is still not fully understood. Klein,
H\"ofener and coworkers \cite{Klein1999,Hofener2000} pointed out
the strong impact of inelastic processes in the barrier region.
They concluded that multi-step tunneling via a number of localized
states within the barrier had to be added to the elastic tunneling
conductivity contribution. Several different inelastic processes
have been suggested in order to explain experimental data, most of
them emphasizing scattering at magnetic intra-barrier states
\cite{Guinea1998,Lee1999,Ziese1999}.

So far, most studies of the transport mechanism in manganite grain
boundaries have focused on the non-linearity of the $I-V$-curves,
the magnitude of the low-field magnetoresistance or the shape of
the magnetoresistance at high fields. The shape of the low-field
\emph{hysteresis}, directly related to the magnetization reversal
process at the grain boundary, has long been neglected. Recently
however, one attempt to reproduce the hysteresis was presented:
Garc\'{\i}a and Alascio \cite{Garcia2002} minimized the total
magnetic energy and obtained the magnetoresistance from the
magnetization directions. They considered the energy within the
grain boundary region, and assumed uniaxial anisotropy in order to
reproduce the experimental data.

In this paper we explore the possibility to describe the low-field
magnetoresistance in a bicrystal junction in terms of coherent
rotation of the magnetization directions, similar to what was
presented more than half a century ago by Stoner and Wohlfart
\cite{Stoner1948}. In a numerical calculation we demonstrate that
the \emph{shape} of the hysteresis in the magnetoresistance can be
obtained by a simple energy-minimization technique, where each
grain is treated individually. We show that a biaxial
magnetocrystalline anisotropy very well reproduces the measured
magnetoresistance curve in a bicrystal grain boundary junction.



\begin{figure}
    \begin{center}
   \includegraphics[width=0.8\columnwidth]{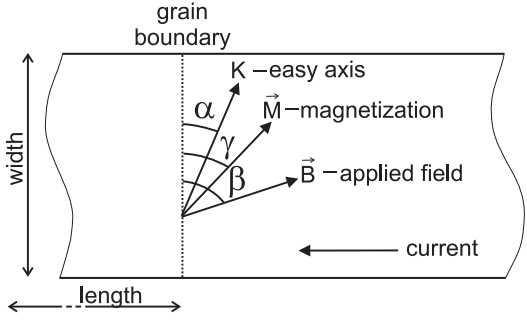}
    \caption{The directions of the easy axis, the applied magnetic field and the magnetization
    ($\alpha$, $\beta$ and $\gamma$, respectively) are all defined with respect to the grain boundary.
    The magnetoresistance is measured with the current along the length of the structure.
    \label{fig:DefAngles}}
    \end{center}
\end{figure}

In the original paper by Stoner and Wohlfart from 1948
\cite{Stoner1948}, they described the magnetization reversal of
uniformly magnetized ellipsoidal particles. Generally the
equilibrium domain structure and magnetization reversal processes
are determined by the balance between the exchange,
magnetocrystalline, magnetostatic and Zeeman energy contributions
(see e.g. Ref.~\cite{Aharoni2000}). Let us consider the
two-dimensional case of a microbridge crossing a single bicrystal
grain boundary, a type of device that has been well studied
experimentally. The easy axis of magnetization, the magnetic field
$\vec{B}$ and the magnetization $\vec{M}$ are oriented in
directions ($\alpha$, $\beta$ and $\gamma$, respectively) defined
relative to the grain boundary as depicted in
Fig.~\ref{fig:DefAngles}. Furthermore, let us first consider the
two single crystal electrode sides (left and right) as being
uniformly magnetized and decoupled from each other, i.e. only the
variation of the magnetocrystalline and Zeeman energies have to be
considered. Hence the field dependent part of the energy density
for one of the sides can be written as
\begin{eqnarray}
\omega=\frac{K}{4} \sin^22(\alpha-\gamma) - M B \cos(\beta -
\gamma), \label{eq:TotalEnergy}
\end{eqnarray}
where $K$ is the (first order) biaxial anisotropy coefficient. In
the case of uniaxial anisotropy the first term should be replaced
by \cite{Note:Landau}
\begin{eqnarray*}
   K_u \sin^2(\alpha-\gamma).
\end{eqnarray*}

\begin{figure}
    \begin{center}
    \includegraphics[width=0.95\columnwidth]{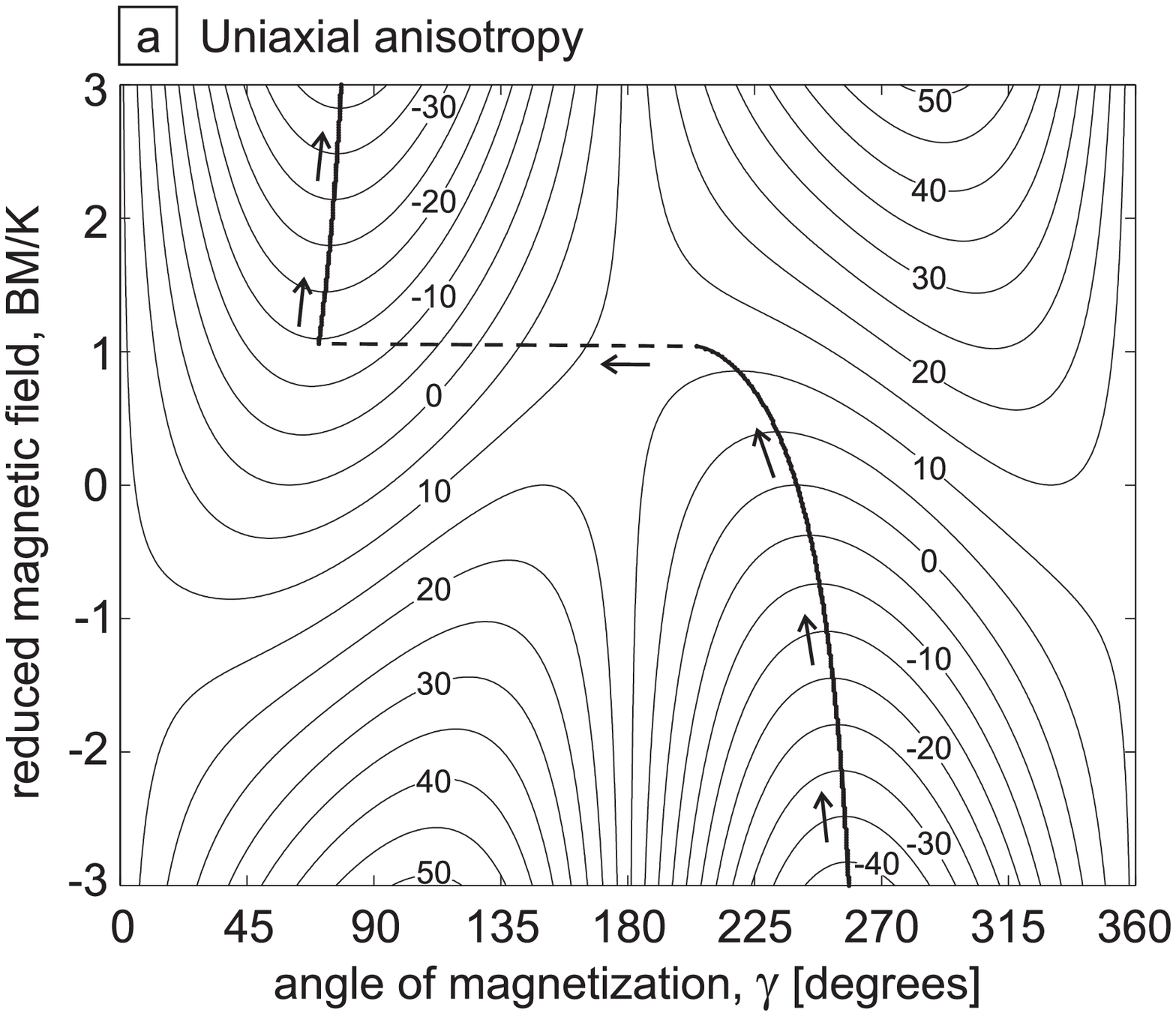}
    \includegraphics[width=0.95\columnwidth]{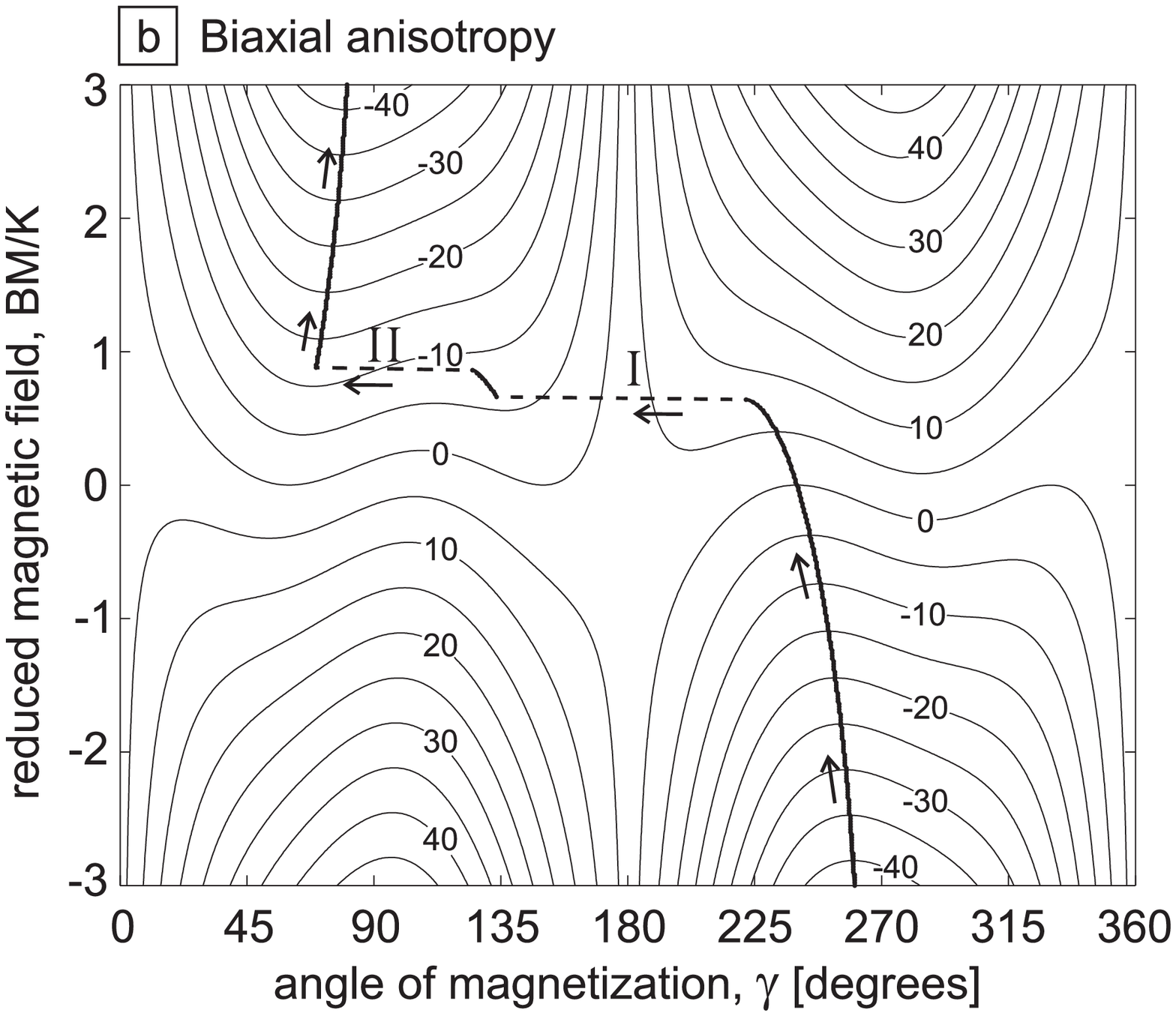}
    \caption{Constant energy contours ($\omega$) calculated from Eq.~\ref{eq:TotalEnergy}
    for constant $K$ and $M$ ($K/M=38$~mT) using a) uniaxial and b) biaxial anisotropy,
    with $\alpha=60^\circ$ and $\beta=90^\circ$. The bold line indicates a local minimum
    traced from $-B_{max}$ to $+B_{max}$, and the direction of rotation is indicated by
    arrows.\label{fig:energylandscape}}
    \end{center}
\end{figure}

For fixed values of $K$ and $M$ and the angles $\alpha$ and
$\beta$, the magnetization direction $\gamma$ can be obtained from
Eq.~\ref{eq:TotalEnergy} by tracing a local energy minimum in the
$\omega(B,\gamma)$ diagram. The starting point in this numerical
method is chosen to be a global minimum at high magnetic field. By
stepping the magnetic field we obtain $\gamma(B)$, as shown in
Fig.~\ref{fig:energylandscape}. With the assumption that
magnetization reversal occurs by coherent rotation and
$|\vec{M}|=M_s$, where $M_s$ is the saturation magnetization, we
can trace the rotation of $\vec{M}$ and find $\vec{M}(B)$.

\begin{figure}
    \begin{center}
    \includegraphics[width=0.85\columnwidth]{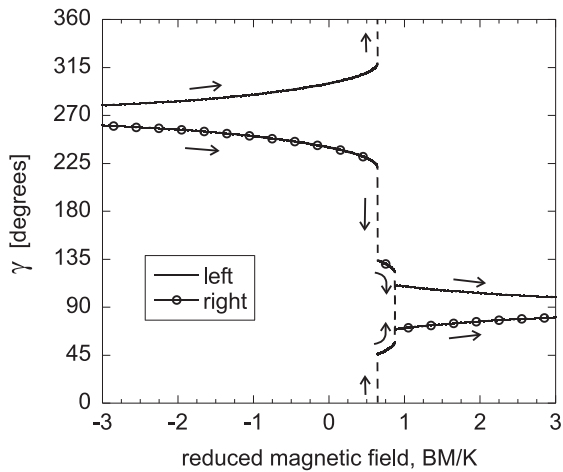}
    \caption{The angle of magnetization $\gamma$ for the left and right electrodes
    with $\alpha_R=-\alpha_L=60^\circ$, $\beta=90^\circ$, and $K/M=38$~mT
    calculated for the case with biaxial anisotropy.
    The magnetic field starts from a high negative field and sweeps to positive values.
    \label{fig:CalcAngles}}
    \end{center}
\end{figure}

To apply this model to the bicrystal grain boundaries we trace the
local minima defined by Eq.~\ref{eq:TotalEnergy} for the left and
right side independently. Hence we obtain $\gamma_L(B)$ and
$\gamma_R(B)$ for the two sides with easy axis in the direction
$\alpha_L$ and $\alpha_R$, respectively. In
Fig.~\ref{fig:CalcAngles} we show $\gamma_L(B)$ and $\gamma_R(B)$
calculated for biaxial anisotropy.

Now, for comparison with transport data, we consider the two sides
connected by a tunneling barrier at the grain boundary. This
yields a ferromagnet/non-magnet/ferromagnet (F/N/F) spin valve
structure. As theoretically demonstrated by Slonczewski
\cite{Slonczewski1989}, the spin-dependent tunneling at the grain
boundary is proportional to the cosine of the angle between the
magnetization directions of the electrodes. Thus the spin
dependent conductivity is
\begin{eqnarray*}
\sigma_{sp} = G_{sp} [1+P^2\cos(\gamma_L-\gamma_R)],
\label{eq:SpinPolarisedConductance}
\end{eqnarray*}
where $G_{sp}$ is the spin polarized conductivity at
$\gamma_L-\gamma_R=\pi/2$, and $P$ is the spin polarization of the
electrodes.  With a non-spinpolarized contribution
$\sigma_{ns}=G_{ns}$, the total resistivity of the tunnel barrier
is
\begin{eqnarray}
\rho = \frac{1}{\sigma_{sp}+\sigma_{ns}} \propto
\frac{1}{1+P^2\cos(\gamma_L-\gamma_R)+G},\label{eq:resistance}
\end{eqnarray}
where $G=G_{ns}/G_{sp}$. Spin-flipping inelastic processes can
also be included in $\sigma_{ns}$. For each value of $B$ we
calculate $\gamma_L$ and $\gamma_R$ and hence obtain the
resistance hysteresis $\rho(B)$.

The results of the numerical simulation are compared to the
measured magnetoresistance of a bicrystal sample. The sample is a
La$_{0.7}$Sr$_{0.3}$MnO$_3$ (LSMO) film grown on a SrTiO$_3$
bicrystal substrate. The bicrystal is a symmetric (001)-tilt grain
boundary with a misorientation angle of $\pm15^\circ$, i.e. the
[100] directions of the left and right sides are in-plane rotated
$15^\circ$ (in opposite directions) with respect to the grain
boundary. The 120~nm thick film was grown by pulsed injection
metal-organic chemical vapour deposition (MOCVD). More details on
the growth procedure can be found in Ref.~\cite{Dubourdieu2001}.
The good epitaxy of the film was verified by x-ray diffraction in
$\theta$--$2\theta$- and $\phi$-scans. The out-of-plane lattice
parameter of the film was found to be 3.86~{\AA}, to be compared
to the bulk pseudo-cubic lattice parameter of the perovskite
structured LSMO of 3.88~{\AA}. It is known that this kind of
tensile strain leads to an in-plane biaxial magnetocrystalline
anisotropy in the $\langle
110\rangle$-directions.\cite{Steenbeck1999,Berndt2000} Hence, this
LSMO on SrTiO$_3$ system can be analyzed with the two-dimensional
model considered for the Stoner-Wohlfart theory. The saturation
magnetization of the film, determined in a superconducting quantum
interference device (SQUID) magnetometer, was $400$~kAm$^{-1}$.

A 5~$\mu$m wide and about 150~$\mu$m long microbridge crossing the
grain boundary was defined by photolithography and Ar-ion milling,
and the ends of the microbridge were connected to electrical
contacts. The resistance of the grain boundary was measured in a
four-contact geometry with a current bias of 10~$\mu$A. The
magnetoresistance ($R(B)/R(0)$) was measured with the field
applied along the length of the microbridge, i.e. in the plane of
the film and perpendicular to the grain boundary.  The magnetic
field was swept with 0.05~T/min. The dependence of the
magnetoresistance on the magnetic field direction is presented in
\cite{Gunnarsson2004}.




The two energy terms in Eq.~\ref{eq:TotalEnergy} are comparable in
size when the field is $B \approx K/M$. For LSMO films on
SrTiO$_3$ numerical values of $K$ of 1.6 --
5.7$\times10^3$~J/m$^3$ \cite{Berndt2000,Steenbeck1999} have
previously been reported. In addition we have $M=400$~kAm$^{-1}$,
as stated above. Hence a simple estimate of the critical field for
magnetization switching would give $K/M=$4 -- 14~mT.

\begin{figure}
    \begin{center}
    \includegraphics[width=0.85\columnwidth]{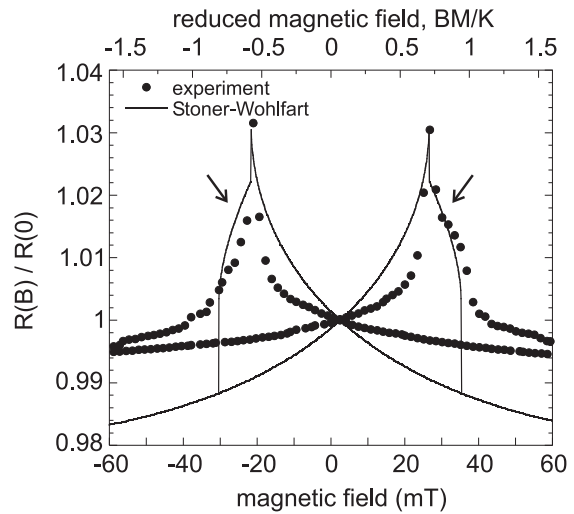}
    \caption{The magnetoresistance of a manganite grain boundary. The circles are measured
    at 1.9~K according to the descriptions in the text. The solid line is a fit to
    Eq.~\ref{eq:resistance} with $P=1$ and $G=20$ (using $K/M=38$~mT in Eq.~\ref{eq:TotalEnergy}).
    The arrows indicate the shoulder discussed in the text.\label{fig:CompareTheoryAndFit}}
    \end{center}
\end{figure}

Since the measured sample has a misorientation angle of
$\pm15^\circ$, and the easy axes are in the [110] and equivalent
in-plane directions, $\alpha_{L,R}=\pm 60^\circ$. The field
applied along the microbridge corresponds to $\beta=90^\circ$.
Using these parameters in Eq.~\ref{eq:TotalEnergy} we can fairly
well reproduce the measured grain boundary magnetoresistance, see
Fig.~\ref{fig:CompareTheoryAndFit}. Here we use a $K/M$ ratio of
38~mT, $P=1$ and $G=20$. Hence, the $K/M$ from the experimental
results is within one order of magnitude of the value estimated
above.

The width of the simulated magnetoresistance hysteresis (the
distance between the peaks) is uniquely determined by the $K/M$
ratio. The general shape of the curve is determined by the chosen
type of magnetocrystalline anisotropy (uniaxial or biaxial)
together with the angles of the easy axes ($\alpha_L$ and
$\alpha_R$). The curvature of the magnetoresistance mainly depends
on the transport model \cite{Note:Hopping}. The height of the
simulated magnetoresistance depends on the values chosen for $P$
and $G$. It is obvious from Eq.~\ref{eq:resistance} that a
decreased spin polarization ($P<1$) has virtually the same effect
as an increased non-spinpolarized current ($G>0$). In the
simulation we used $P=1$ and $G=20$, but for instance $G=0$ and
$P=0.22$ would yield the same magnetoresistance magnitude. Without
further information about either $P$ or $G$, a ratio between the
two can not be extracted from the measured data with our model.


We note that in the measured curve, there is a peak and a shoulder
on each wing of the hysteretic curve. Both these features are
represented in the curve (solid line in
Fig.~\ref{fig:CompareTheoryAndFit}) calculated from
Eq.~\ref{eq:resistance}. Each feature is associated with a jump in
the magnetization direction. A closer look into the model of
biaxial anisotropy reveals that the first jump comes from the
switching from a high energy state to one with lower energy,
marked \textrm{I} in Fig.~\ref{fig:energylandscape}b. In the
low-energy valley there are two local minima. Thus the first part
of the magnetization rotation is paused when the first of the two
minima is reached in the direction of rotation. However, under
some geometrical conditions there exists a state even lower in
energy, and the second jump (\textrm{II}) is a transition to that
state. The dwelling in the intermediate state creates the shoulder
(marked by arrows in Fig.~\ref{fig:CompareTheoryAndFit}) in the
magnetoresistance curve. We note that this intermediate state does
not exist in the case of uniaxial anisotropy (see
Fig.~\ref{fig:energylandscape}a). Thus, we conclude that with the
Stoner-Wohlfart model we can well reproduce the characteristic
features of the low-field magnetoresistance related to bicrystal
grain boundaries, in the case when the magnetic field is
perpendicular to the grain boundaries and a biaxial anisotropy is
included.


We find the agreement between the numerical values obtained in the
model and in experiments to be good, although there are details
that are not fully reproduced. In the model, the
magnetocrystalline anisotropy $K$ and the magnetization $M$
uniquely determine the position of the peak. Considering the
uncertainty in the experimental material parameters the estimated
position agrees well with the experiments. By including surface
effects at the grain boundary, which may locally decrease $M$ and
enhance the effective value of $K$, the agreement can be improved.

Our simulation underestimates the curvature of the
magnetoresistance curve. The curvature is determined by the
transport model and here we have used the one by Slonczewski
\cite{Slonczewski1989}, who considered a single-band direct
tunneling process. From $I-V$ measurements we know that direct
tunneling is insufficient to explain the transport properties in
this kind of junctions \cite{Klein1999}. We also note that an
enhanced curvature has been observed with an increased bias
current \cite{Westerburg1999}. Hence, our results propose that a
more complete transport model for this kind of magnetic tunnel
junctions should be developed. Furthermore, the good agreement
between the Stoner-Wohlfart model and the shape of the
magnetoresistance curves motivates future studies of the
electrical transport in this kind of systems.

The presence of the shoulder in the magnetoresistance hysteresis
and its relation to the biaxial anisotropy have not, to our
knowledge, been explicitly studied previously. Garc\'{\i}a and
Alascio \cite{Garcia2002} did use a Stoner-Wohlfart-like approach
to the problem, but failed to demonstrate the influence of biaxial
anisotropy. However, Philipp \textit{et al} \cite{Philipp2000} and
Todd \textit{et al} \cite{Todd2001} present magnetoresistance
curves for single magnetic bicrystal junctions with the field
applied perpendicular to the grain boundary, and both curves show
a behaviour similar to what can be expected from a coherent
rotation of magnetisation direction with biaxial
magnetocrystalline anisotropy.

A reason for the shortage of experimental data may be that the
shoulder only appears under two conditions: at low temperatures
and for a certain range of angular relations between the grain
boundary, the anisotropy axes, and the applied magnetic field. In
the paper by Gunnarsson et al. \cite{Gunnarsson2004}, which shows
measurements on the same sample as presented here, the shoulder is
absent in the magnetoresistance hysteresis measured at 100~K (with
the field perpendicular to the grain boundary). One reason for the
$T$-dependence is of course the strong variation of the magnetic
properties with temperature. It should be noted that the
Stoner-Wohlfart model is valid for coherent rotation and thus
sufficient to describe the case when the field is applied
perpendicular to the grain boundary, in our experimental setup.
With the field applied in another direction (e.g. parallel to the
grain boundary \cite{Gunnarsson2004}) the magnetization reversal
processes become more complex. In that case one has to include
contributions from the magnetostatic and exchange energy terms in
the simulation.


In summary, we proposed that the Stoner-Wohlfart theory can be
applied to manganite grain boundary magnetoresistance. Our
conclusion is that the model of two magnetically decoupled
electrodes with coherently rotating magnetization vectors very
well explain the experimental data. We find that in the case of
La$_{0.7}$Sr$_{0.3}$MnO$_3$ junctions on SrTiO$_3$ bicrystals it
is necessary to include a biaxial magnetocrystalline anisotropy to
fully reproduce the measured data. In addition our study shows
that a bicrystal grain boundary, with its well defined angles of
magnetization, constitutes a system well designed to explore the
physics of magnetic tunnel junctions.

\acknowledgments We are grateful for the financial support from
Vetenskapsr{\aa}det (VR) and Stiftelsen f\"{o}r Strategisk
Forskning (SSF).



\begin{thebibliography}{25}
\bibitem[{Not({\natexlab{b}})}]{Note:Hopping}
\bibinfo{note}{With a double exchange, which is a spin-polarized hopping
  conductivity where $\sigma_{sp}\propto\cos^2(\gamma_L-\gamma_R)$, the
  curvature is even less, and hence the agreement with experiment is worse.
  This transport model was used by Garc\'{\i}a and Alascio \cite{Garcia2002}.}
\end{thebibliography}
\end{document}